\documentstyle[12pt]{article}

\begin{document}
\title{Theories for multiple resonances}
\author{D. Klakow, M. Weber, and P.-G. Reinhard \\
  Institute f\"ur Theoretische Physik II,  Staudtstr. 7 \\
  Universit\"at Erlangen-N\"urnberg, D-91058 Erlangen, Germany}
\date{\today}
\maketitle

\begin{abstract}
Two microscopic theories for multiple resonances in nuclei are compared,
n-particle-hole RPA and quantized Time-Dependent Hartree-Fock (TDHF).
The Lipkin-Meshkov-Glick model is used as test case.
We find that quantized TDHF is superior in many respects, except
for very small systems.
\end{abstract}

\section{ Introduction}

The most prominent excitations of the nucleus are the giant resonances.
One has therefore dreamed about multiple excitation of giant resonances
quite early in the development of nuclear theory \cite{Brink}. Recent
experimental progress has brought those multiple resonances into the
range of observability \cite{expDR}. This has inspired several
theoretical investigations in the framework of a second RPA, see e.g.
\cite{Yan,Chomaz,Lauritsch}. Multiple excitations of basic collective
modes are well known from the low energy-spectra of soft nuclei.
They are most often characterized phenomenologically in terms of
interacting elementary Bosons, e.g. surface vibrations and pairing modes
\cite{IBM}.  Theoretical foundations for such a collective description
of the low-energy modes have been long searched for in the framework of
the Boson expansions, for a review see \cite{BosExp}. This
task has turned out to be very involved, and a possible explanation is
that the low energy modes are far from a simple picture of collective
motion as mere shape vibrations of the mean field. Motion along a
deformation path is accompanied by dramatic reoccupations amongst
crossing single particle levels \cite{ReiOtt} having possibly curious
consequences for the collective dynamics \cite{BulgVort}. Giant
resonances, on the other hand, represent the much cleaner collective
modes, dominated by the mean field and without noticeable pairing
effects. The theoretical appeal of the multiple resonances is then to
study large amplitude motion in a case where the underlying elementary
Boson is fairly well understood. The basic modes, the giant resonances,
are usually described within the RPA, the theory of small $1ph$
excitation about the Hartree-Fock ground state \cite{Rowe1}. Multiple
resonances require extensions beyond standard RPA. A straightforward
next step is the "second RPA" extending the treatment to the space of
$2$-$ph$ excitations \cite{Yan,Lauritsch}. Boson expansions, which have
also been considered in that context \cite{Chomaz} confine the extension
to the collective modes but allow to go further in the anharmonic terms.
Somewhat more general are the "higher RPA" approaches considering
coherent mixtures of $n$-$ph$-excitations. The equations-of-motion
techniques as outlined in \cite{Rowe1,Rowe} provide an extremely useful
tool to derive those expansions along a given set of $n$-$ph$ operators.
A much different point of view is presented by Time-Dependent
Hartree-Fock (TDHF) approach which pronounces more the vibrating mean
field. Its small amplitude limit delivers also the RPA. The spectra for
higher modes can be computed with quantized TDHF which picks the quantum
states amongst all the classically allowed energies by requiring
strictly periodic TDHF orbit which have integer action along
one closed orbit \cite{Negele}. It is the aim of this paper to compare
the results on multiple resonances as computed with quantized TDHF or
with an $n$-$ph$-RPA derived from an equations of motion technique. The
comparison is performed in a schematic model for nuclear dynamics in the
active shell around the Fermi surface, the Lipkin-Meshkov-Glick (LMG)
model \cite{Lipkin,Meshkov}. We will use in the following the
abbreviation RPA-$N$ for such a higher RPA which includes up to
$N$-$ph$ excitations.

The paper is outlined as follows:
In section~\ref{secLip}, we introduce briefly the LMG model.
In section~\ref{secEoM}, we recapitulate the equations-of-motion
technique and work it out for the present model.
In section~\ref{secTDHF}, we explain quantized TDHF and how we
compute it.
Finally in section~\ref{secRes}, we present and discuss the results
from both schemes.

\section{The Lipkin Model}
\label{secLip}

The LMG model \cite{Lipkin,Meshkov} is a simplified shell model with two
degenerate bands of single Fermion states $a^\dagger_{s,m}$, $s\in\{+,-\}$,
$m\in
[1...\Omega]$. The uncoupled ground state of the model carries all
states $s=-$ as occupied and $s=+$ as empty. Due to the degeneracy, all
single particle excitation energies are the same $\epsilon$. The model
Hamiltonian reads
\begin{equation}
  H=\epsilon K_0 + \chi\frac{\epsilon}{2(\Omega-1)}(K_+ K_+ +K_- K_-)
\label{LMGHam}
\end{equation}
where the basic operators are
\begin{eqnarray}
  K_0 &=& \sum_{m=1...\Omega} \left[ a^\dagger_{+,m}a^{\mbox{}}_{+,m}
                                  -a^\dagger_{-,m}a^{\mbox{}}_{-,m} \right]
\nonumber
\\
  K_+ &=& \sum_{m=1...\Omega}  a^\dagger_{+,m}a^{\mbox{}}_{-,m}
\\
  K^{\mbox{}}_- &=& K_+^\dagger
\end{eqnarray}
The second term, quadratic in the $K_\pm$, models the residual
interaction. The $K_\pm$ are $1ph$ operators with respect the uncoupled
ground state and thus the Hamiltonian (\ref{LMGHam}) is tailored for
model studies of the dynamics along the $1ph$ channel, i.e. for mean
field dynamics. The dimensionless parameter $\chi$ describes the
interaction strength to the single=particle splitting $\epsilon$. The
scaling with $1/(\Omega-1)$ places the critical coupling at $\chi=1$
almost independent of the system size $\Omega$.

Note the two important features of the LMG model: First, the residual
interaction is only active for vertical $1ph$ excitations, i.e.
those which leave the secondary quantum number $m$ invariant.
Second, all substates $m$ are handled with the same weight and phase
in each of the three operators $K_\alpha$. This introduces a high symmetry
in the model such that we have an exactly decoupling collective
subspace which is spanned by the Thouless transformations
$|\Phi\rangle = \exp{\left(i\sum_{\alpha\in\{-,0,+\}}c_\alpha K_\alpha\right)}
|0\rangle$ with the basic collective operators $K_\alpha$ \cite{Thouless}.
Moreover, these basis operators form a simple $SU(2)$ algebra
\begin{equation}
  [K_0,K_\pm]=\pm K_\pm \qquad {\rm and } \qquad [K_+,K_-]= 2 K_0
  \qquad .
\label{SU2}
\end{equation}
All further evaluations, in the equations-of-motion technique
as well as in quantized TDHF, require only this algebra and, of
course, the model Hamiltonian (\ref{LMGHam}).

In spite of its simplicity, the LMG model unfolds a rich variety of
scenarios for the collective motion, nearly harmonic motion for small
$\chi$, increasing anharmonicities with increasing $\chi$, and unstable
motion with subsequent phase transition near $\chi=1$, for a
visualization see the beginning of section~\ref{secRes}. It provides the
ideal testing ground for our purposes. In fact, it was one of the
originals aims of the LMG model to understand the appearance of
collective modes in the RPA. It is the mode which is excited by a mix
$xK_+-yK_-$ and whose frequency is $\omega=\epsilon\sqrt{1-\chi^2}$ in
the domain $\chi<1$.

\section{Equations-of-motion and RPA-$N$}
\label{secEoM}

The equations-of-motion technique as outlined in \cite{Rowe,Rowe1} is in
our opinion the most obvious way to formulate a RPA-$N$ as an algebra of
selected excitation operators. The aim is to optimize a set of
excitations operators $Q^{\dagger}_\nu$ which excite the state
\begin{equation}
  |\nu\rangle=Q^{\dagger}_{\nu} |0\rangle
\label{exstate}
\end{equation}
out of a correlated ground state $|0\rangle$.
The excitation operators are taken as linear superposition from
a basis set of operators ${\cal Q}_\alpha$
\begin{equation}
  Q_{\nu} = \sum_\alpha c_\alpha^\nu {\cal Q}_\alpha
  \qquad.
\label{superp}
\end{equation}
In the ideal case, the excitation operators should fulfill he
Heisenberg equation
\begin{equation}
   [H,Q^{\dagger}_\nu] = \left(E_{\nu}-E_0\right)Q^{\dagger}_\nu
   \qquad.
\label{Opeq}
\end{equation}
But this can rarely be fulfilled in practice because the algebra
of the ${\cal Q}_\alpha$ and $H$ is not closed. One thus requires
Eq.~(\ref{Opeq}) in the average after "stabilizing" the expression
to a double commutator. This leads to the variational equations
\begin{equation}
  <0|[\delta Q,[H,Q^{\dagger}_{\nu}]]|0>
  =\omega_\nu<0|[\delta Q ,Q^{\dagger}_{\nu}]|\nu>
  \quad,\quad
  \omega_\nu=(E_{\nu}-E_0)
\label{RPA}
\end{equation}
where $\delta Q$ stands for any variation in the given space of ${\cal
Q}_\alpha$. The correlated ground state $|0\rangle$ can be generated
with a similar variational equation. It is written in the form
\begin{equation}
  |0>=e^S |HF>
\end{equation}
where $S$ is an antihermitian operator and $|HF>$ the Hartree-Fock
ground state. The Operator $S$ is determined by \cite{Rowe}
\begin{equation}
  <0| [\delta S,H ] |0> = 0
\label{gr1}
\end{equation}
where $\delta S$ is again any variation in the given basis space
of operators.

The spectrum of collective excitations in the Lipkin Model is generated
from the basis states $K_+^n$ and $K_-^n$. We thus make for the
excitation operator the ansatz
\begin{equation}\label{exop}
  Q_\nu^{\dagger} = \sum_{n=1}^{N} x_{n,\nu} K_+^n - y_{n,\nu} K_-^n
\end{equation}
and similarly for the correlation operator
\begin{equation}
  S=\sum_{n=1}^{M} \gamma_{n,\nu} \left(K_+^n -  K_-^n\right)
  \qquad.
\end{equation}
The $N$ and $M$ represent the order of the expansion, in practice the
maximal order of $N$-$ph$ or $M$-$ph$ operator taken into account. The
equations-of-motion (\ref{RPA}) and (\ref{gr1}) constitute a set of
equations to determine $\omega_\nu$, $x_{i,\nu}$, $y_{i,\nu}$ ($n=1 ..
N\,,\,\nu=1 .. N$) and $\gamma_n$ ($n=1..M$). It is most consistent to
use the same order $M\approx N$ in both pieces and this yields usually
the best results for a given expense. An exception is the conventional
RPA which appears as the case $N=1$ and $M=0$ within that scheme. The
enhancement to $M=1$ is unnecessary in that case because the $1ph$
states decouple from the Hartree-Fock ground state \cite{Ring}. For any
other $N>1$, it is best to chose $N=M$ and this is what we call the
higher RPA of order $N$, in short RPA-$N$. We will consider in the
following the two particular examples RPA-$2$ and RPA-$5$.

\section{Quantized TDHF}
\label{secTDHF}

TDHF approximates the dynamics of a many-Fermion system by a
time-dependent Slater state $|\Phi(t)\rangle$. The time-evolution of
$|\Phi(t)\rangle$ is optimized by deriving the TDHF-equations from the
variational principle of stationary action
\begin{equation}
  \delta\int <\Phi|i\hbar{d\over dt}-H|\Phi>\,dt = 0
\label{varp}
\end{equation}
where $\delta$ denotes variations within the subspace of Slater states.
TDHF delivers a deterministic equation-of-motion for $|\Phi(t)\rangle$.
It can be interpreted as representing the classical limit of
many-Fermion dynamics. The interpretation is pertinent in particular
with respect to the fact that TDHF motion is possible for any initial
condition $|\Phi(0)\rangle$ at any energy. However, it can be
complemented by a (semiclassical) quantization, as has been shown by
very different approaches, as e.g. the functional integral
representation of the Green function $G(E)={\rm tr}{1\over \hat H-E}$
\cite{Levit,Blaizot,Reinhardt} or the method of gauge invariant periodic
quantization \cite{Kan,Kan1}. The quantum states are those solutions of
the TDHF equation (\ref{varp}) which fulfill two additional conditions:
First, they are strictly periodic including the phase,
\begin{equation}
  |\Phi(t+T)\rangle = |\Phi(t)\rangle
  \qquad,
\label{percond}
\end{equation}
and second, they fulfill the quantization condition
\begin{equation}
  \int_0^T<\Phi|i\hbar{d\over dt} |\Phi> = 2n\pi \hbar
\label{quancond}
\end{equation}
where $n$ is some integer number.

The most problematic part of quantized TDHF is to find periodic
solutions \cite{Baranger} in the chaotic manifold of trajectories of a
realistic TDHF calculation. Fortunately, this is no problem at all in
the LMG model because the motion of interest is reduced to the one
collective degree-of-freedom spanned by the $K_\pm$ as defined in
section~\ref{secLip}. First tests of quantized TDHF have therefore been
performed within this model \cite{Levit,Kan}. According to the Thouless
theorem, each Slater determinant of a given system can be generated from
a reference determinant $|\Phi_0\rangle$ as
$|\Phi\rangle=\exp{(A_{ph})}|\Phi_0\rangle$ where $A_{ph}$ is a $1ph$
operator. We aim to span the space of "collective deformations" outgoing
from the Hartree-Fock ground state. This is achieved by the coherent
states
\begin{equation}
  |z> = e^{zK_+}|HF>(1+|z|^2)^{-{\Omega\over 2}}
\label{state}
\end{equation}
which are labelled by the complex shift parameter $z$. Note that these
coherent states are already normalized.

The expectation values which are necessary for Eq.~(\ref{varp})
are easily evaluated using the algebra (\ref{SU2}) and the subsequent
simple properties of the coherent states. We obtain
\begin{eqnarray}
  <z|i\hbar{d\over dt}|z> &=& -{\Omega \over 2}\hbar (1-\cos \psi)\dot\phi
\label{dtexp}
\\
  <z|H|z> &=&
   -{\Omega\epsilon \over 2}(\cos \psi -{\chi \over 2}\sin^2\psi\cos2\phi )
\label{Hexp}
\\
  z &=& e^{i\phi}\tan{\left(\frac{\psi}{2}\right)}
\label{ztrans}
\end{eqnarray}
The time evolution of the coherent state is contained in the time-evolution
of the parameter $z$.
By Variation according to (\ref{varp}) we find the TDHF equations
as equations-of-motion for $z$
\begin{eqnarray}
  {\hbar \over \epsilon}\dot\psi &=& -\chi \sin\psi\sin 2\phi
\label{psidot}
\\
  {\hbar \over \epsilon}\dot\phi &=& -(\chi\cos\psi\cos 2\phi+1)
\label{phidot}
\end{eqnarray}
The periodicity condition (\ref{percond}) is trivially fulfilled because
we have an effectively two-dimensional phase space. It remains to watch
the quantization condition (\ref{quancond}) which reads in the LMG model
\begin{equation}
  -{\Omega\over 2}\int^T_0(1-\cos\psi)\dot\phi\, dt
  = -{\Omega\over 2}\oint (1-\cos\psi)\, d\phi
  = 2n\pi
  \qquad.
\label{quantLMG}
\end{equation}
The initial condition of the coupled differential equations
(\ref{psidot}) and (\ref{phidot}) are varied until the solution fulfills
the quantization condition (\ref{quantLMG}). The (quantum) excitation
energy is then provided by the expectation value (\ref{Hexp})
which is constant along the stationary path.

\section{Results}
\label{secRes}

The LMG model has essentially two parameters, namely $\chi$, the
relative coupling strength, and $\Omega$, the particle number or size of
active phase space. Variation of $\chi$ varies the ground state
deformation and the anharmonicity of the collective resonances. The
uncoupled ground state of the model is the Hartree-Fock ground state for
$\chi<1$ whereas a transition to a "deformed" state occurs for $\chi>1$.
Close to harmonic excitations appear at $\chi\ll 1$ and $\chi\gg 1$
whereas increasing anharmonicities build up if the critical point
$\chi=1$ is approached from both sides. Variation of $\Omega$ influences
the softness of the transition around the critical $\chi=1$, being
softer for small $\Omega$ and developing a sharp phase transition for
$\Omega\longrightarrow\infty$. In order to visualize the dynamics of the
system, we show in Fig.~\ref{PES} the deformation energies versus
collective deformation angle $\psi$.

The case $\chi=0.4$ displays a nearly parabolic potential indicating
that we have to expect essentially harmonic motion there with small
anharmonic perturbations. The case $\chi=1$ shows an extremely soft and
anharmonic potential. The system is in a critical regime where small
perturbations lead to huge reactions. The effect remains finite for
finite $\Omega$ and tends to critical, i.e. infinite, fluctuations for
$\Omega\longrightarrow\infty$. The case $\chi=4$ shows again a well
developed ground state with small oscillations about it in the first
excited state. But note that this ground state is placed at finite
deformation $\psi$ and that it appears twice at exactly the same energy.
As a consequence, we will have a doubled excitation spectrum because
each mode has a symmetric copy about the other ground state. The
excitation energies from quantized TDHF or RPA-$N$ are exactly
degenerate. The degeneracy of the exact excited states is removed a
little bit due to tunneling processes through the barrier at $\psi=0$.
None of the presently discussed theories can reproduce this splitting.

In Fig.~\ref{softchi}, we compare the results of quantized TDHF, RPA-$5$
and RPA-$2$ with the exact solution up the fourth resonance excitation.
The figure concentrates on the regime of undercritical couplings. Three
system sizes are considered which are comparable in the nuclear case to
the phase spaces of light, medium and heavy nuclei. The first excited
state is very well described by all methods. The tiny differences seen
there would favour quantized TDHF and, somewhat surprisingly, RPA-$5$
looks a bit inferior to RPA-$2$ for small $\Omega$. But one should not
overstress these details. The comforting message is that plain RPA
performs well for the first excited state in the regime of undercritical
coupling. More visible differences show up with increasing excitations.
Quantized TDHF looks inferior for the small system size $\Omega=5$ where
RPA-$5$ performs very well. The reverse happens for medium and large
systems where the result from quantized TDHF are astonishingly close to
the exact solution. This complies with the interpretation of TDHF as a
classical limit of the many-Fermion dynamics. This is a classical limit
in the sense of a $1/\Omega$ expansion and it becomes increasingly valid
with increasing system size, here $\Omega$. On the other hand, the
success of the RPA-$5$ for $\Omega=5$ is not so surprising because that
expansion is then close to complete (note that it is not yet fully
complete at 5th order because the $0$-$ph$ operator $K_0$ is missing in
the expansion and has to be regenerated from pairs of $K_\pm$). Finally,
we note that the RPA-$2$ can by definition only describe the system up
to the second excited state. But it performs surprisingly well at its
upper limit $\omega_2$ and can reliably be used instead of RPA-$5$
there.

In Fig.~\ref{strongchi}, we show the same comparison for a wider range
of couplings $\chi$. The exact spectra reflect nicely the appearance of
two equivalent ground states in that pairs of excited states develop for
$\chi\longrightarrow\infty$. The first excited state at $\chi<1$ goes to
zero excitation and thus merges into one of the two ground states of the
system, in fact it represents the antisymmetric combination whereas the
former ground state develops into the symmetric combination. The
transition is best developed for the large $\Omega$ whereas the small
$\Omega=5$ stays far from the asymptotic stage within the range of
$\chi$ considered. It is interesting to see that the different
approximations handle the transitional region much differently. The
plain RPA is known to break down near the critical $\chi$
\cite{Holzwarth}. This feature persists also for the higher RPA-$N$.
Even the RPA-$5$ runs into difficulties if $\chi$ grows near $1$. This
region has to be excluded from any RPA-like expansion. It is a typical
application case for theories of large amplitude collective motion, as
e.g. the generator-coordinate method \cite{Holzwarth}. RPA becomes
applicable again beyond the transition where well bound ground state
minima develop again, see Fig.~\ref{PES}. Accordingly, a new branch of
RPA solutions can be computed. We show in Fig.~\ref{strongchi} the
results from the RPA-$2$ because we have there at most a second excited
state and we know from Fig.~\ref{softchi} that this level of
approximation is fully sufficient up the second excitation. The RPA
solutions in the regime of large $\chi$ look fair for large $\chi$ but
run also in insurmountable difficulties if $\chi$ is lowered towards the
critical $\chi=1$. We thus have a separate branch of RPA solutions for
large $\chi$ which is disconnected from the branch of RPA solutions in
the regime of small couplings. As expected from Fig.~\ref{PES}, the
motion in the deformed minima is much less harmonic. As a consequence,
the RPA-$2$ gives less reliable results in the regime $\chi>1$, in
particular for the small system. For example, it misses the coupling
through the barrier and gives always exactly degenerate pairs of
excitation states.

It is interesting to see that quantized TDHF is much more robust at the
critical point and provides surprisingly appropriate results there. It
is indeed able to stretch branches of solutions over all ranges of the
coupling $\chi$. However, a bit of care is still required because once
in a while a branch changes discontinously. This happens in the strong
coupling regime whenever the quantized energy crosses the barrier
between the left and the right minimum. TDHF as a classical theory
cannot describe tunneling through the barrier. Thus the branch of
extended motion over the barrier ceases to exist and two new branches
below the barrier develop in the separate minima. To give an example,
the first excited TDHF state in case of $\Omega=5$ stops at $\chi
\approx 3$. This reflects the fact that it should merge into the second
ground state of the system, see the exact solution for comparison. Thus
the dashed line in the figure should thought to be discontinously
connected to $\omega_1=0$ and this behaviour serves as an approximation
to the exact curve, a crude approximation indeed, but at least the
behaviour is qualitatively present. Similar considerations apply in all
the other cases where a TDHF branch stops. We thus see that quantized
TDHF gives a very robust description in all ranges of $\chi$. It is
quantitatively correct almost everywhere, except at the discontinuities
of the branches. It provides at least the proper qualitative behaviour
in those critical cases where branches of solutions disappear or creep
up. And, these critical cases can be clearly identified such that one
knows where to trust the solution and where better not.

\section{Conclusion}

In this paper, we have investigated two microscopic theories for
multiple resonance excitations in nuclei, the RPA-$N$, an extension of
the RPA to higher $n$-$ph$ states, and quantized TDHF. The RPA-$N$
stands for the class of theories which are based on opertor expanions
and the Heisenberg equations-of-motion. A close relative to RPA-$N$ are
the Boson Expansions such that our findings apply for these as well. The
RPA-$N$ is an expansion in orders of excitations which is easiest and
most reliable for low excitation modes and becomes increasingly
cumbersome with increasing energies and increasing particle number. The
quantized TDHF, on the other hand, concentrates on the time-dependent
state of the system in the Schr\"odinger picture. The mean-field
approximation, employed in TDHF, corresponds to the classical limit of
the dynamics of the many-body system. It thus becomes increasingly valid
with increasing energy and increasing particle number. From this formal
point-of-view, RPA-$N$ and quantized TDHF cover contrary regimes.
RPA-$N$ starts an expansion from the lowest excitation modes and stays
fully quantum mechanical throughout each step whereas quantized TDHF as
a classical approach looks more appropriate for high energies and has to
be semiclassically requantized.

The practical comparison has been performed within the
Lipkin-Meshkov-Glick (LMG) model which mimics the nuclear excitation
dynamics as well as a structural transition from symmetric to deformed
ground states. We have studied three typical cases representing a small,
a medium, and a heavy nucleus. The RPA-$N$ is found to work fine in the
domain of well bound systems, i.e. far below and far above the critical
transition point. However, it runs into serious difficulties around the
critical point which cannot be surmounted even by high order expansion
(actually we have tested it up to $N$=$5$). Quantized TDHF, on the other
hand, provides surprisingly robust results for each excitation energy
and in each structural domain including the transition point. The
classical character of TDHF shines through in the fact that it is
somewhat less reliable for the smallest system in our sample. We thus
conclude that quantized TDHF is superior in most domains, but RPA-$N$
remains the method of choice for small systems at not too high
excitation energies.

The simplicity of the LMG model inhibits practical estimates of the
expense of the methods. The RPA-$N$ at full extend will become
unfeasible for a realistic nucleus, except perhaps $N=2$ for small
nuclei. Higher $N$ can be treated as Boson Expansion with a basis of
fixed $1$-$ph$ excitation operators. The necessary algebra of multiple
commutators can be carried through nowadays but will become very
cumbersome. Quantized TDHF, on the other hand, is formally very simple
as it employs always as a unique ingredient a TDHF code which is nowadys
easily available. But it poses the serious technical problem to find the
branches of periodic orbits in the chaotic multitude of solutions. This
could turn out to be a big hindrance for the application of the
otherwise very straightforward and robust approach.

\bigskip

We would like to thank C. Toepffer for his constant interest on the work
and many helpful comments. D.K. and M.W. have been supported by the
Studienstiftung des Deutschen Volkes.

\newpage

\begin{figure}
\caption{\label{PES}}
{\sl Visualization of the deformation energy (solid lines) versus the
   collective deformation angle $\psi$ for
   the LMG model with $\Omega=14$ particles at three different coupling
   strengths as indicated. The energy of the first excited mode is
   indicated by a dashed line in each case.
    }
\end{figure}

\begin{figure}
\caption{\label{softchi}}
{\sl Spectra of the LMG model as a function of the strength of the
   residual interaction in the range of undercritical coupling $\chi<1$
   for three particle numbers $\Omega$ as indicated. The different line
   types represent different approximation schemes:
   solid       $=$ exact solution,
   dashed      $=$ quantized TDHF,
   dash-dotted $=$ RPA-$5$,
   dotted      $=$ RPA-$2$.
  }
\end{figure}

\begin{figure}
\caption{\label{strongchi}}
{\sl Spectra of the LMG model as a function of the strength of the
   residual interaction in a broad range of couplings for three
   particles number $\Omega$as indicated. The different line types
   represent different approximation schemes:
   solid       $=$ exact solution,
   dashed      $=$ quantized TDHF,
   dash-dotted $=$ RPA-$5$, 'spherical' ground state and RPA-$2$, 'deformed'
ground states,
   dotted      $=$ RPA-$2$, 'spherical' ground state.
Note that a RPA-$5$ on a 'spherical' ground state and a RPA-$2$, 'deformed'
ground states both describe the first six eigenstates of the Hamiltonian.
  }
\end{figure}

\end{document}